\documentclass[aps,pra,twocolumn,superscriptaddress,showpacs]{revtex4}

\usepackage{graphicx}% Include figure files
\usepackage{epsfig}
\usepackage[english]{babel}
\usepackage{amsbsy}
\makeatother

\def\lsim{\mathrel{\rlap{\lower4pt\hbox{\hskip1pt$\sim$}}
    \raise1pt\hbox{$<$}}}                % less than or approx. symbol
\def\gsim{\mathrel{\rlap{\lower4pt\hbox{\hskip1pt$\sim$}}
    \raise1pt\hbox{$>$}}}                % greater than or approx. symbol
\begin{document}

%%\title{ Matter reflects Antimatter }
\title{ Experimental evidence of antiproton reflection by a solid surface }

\author{A. Bianconi}
\affiliation{Dipartimento di Chimica e Fisica per l'Ingegneria e per i Materiali,
 Universit\`a di Brescia, 25133 Brescia, Italy}
\affiliation{Istituto Nazionale di Fisica Nucleare, Gruppo Collegato di Brescia, 25133 Brescia, Italy}

\author{M. Corradini}
\affiliation{Dipartimento di Chimica e Fisica per l'Ingegneria e per i Materiali,
 Universit\`a di Brescia, 25133 Brescia, Italy}
\affiliation{Istituto Nazionale di Fisica Nucleare, Gruppo Collegato di Brescia, 25133 Brescia, Italy}

\author{A. Cristiano}
\affiliation{Dipartimento di Chimica e Fisica per l'Ingegneria e per i Materiali,
 Universit\`a di Brescia, 25133 Brescia, Italy}
%%%\affiliation{Istituto Nazionale di Fisica Nucleare, Gruppo Collegato di Brescia, 25133 Brescia, Italy}

\author{M. Leali}
\affiliation{Dipartimento di Chimica e Fisica per l'Ingegneria e per i Materiali,
 Universit\`a di Brescia, 25133 Brescia, Italy}
\affiliation{Istituto Nazionale di Fisica Nucleare, Gruppo Collegato di Brescia, 25133 Brescia, Italy}

\author{E.~Lodi Rizzini}
\affiliation{Dipartimento di Chimica e Fisica per l'Ingegneria e per i Materiali,
 Universit\`a di Brescia, 25133 Brescia, Italy}
\affiliation{Istituto Nazionale di Fisica Nucleare, Gruppo Collegato di Brescia, 25133 Brescia, Italy}

\author{L. Venturelli}
\affiliation{Dipartimento di Chimica e Fisica per l'Ingegneria e per i Materiali,
 Universit\`a di Brescia, 25133 Brescia, Italy}
\affiliation{Istituto Nazionale di Fisica Nucleare, Gruppo Collegato di Brescia, 25133 Brescia, Italy}

\author{N. Zurlo}
\affiliation{Dipartimento di Chimica e Fisica per l'Ingegneria e per i Materiali,
 Universit\`a di Brescia, 25133 Brescia, Italy}
\affiliation{Istituto Nazionale di Fisica Nucleare, Gruppo Collegato di Brescia, 25133 Brescia, Italy}

\author{R. Don\`a}
\affiliation{Dipartimento di Fisica dell'Universit\`a di Bologna 
and INFN Sezione di Bologna, 40127 Bologna, Italy}

%\thanks{}
\affiliation{}

\date{\today}

\begin{abstract}
We report here an experimental evidence of the reflection  
of a large fraction of a beam of low energy antiprotons by an aluminum wall.
This derives from the analysis of a set of annihilations of antiprotons
that come to rest in rarefied helium gas after hitting the end wall of the apparatus. 
A Monte Carlo simulation of the antiproton path in aluminum indicates that the 
observed reflection 
occurs primarily via a multiple Rutherford-style scattering on Al nuclei, at 
least in the energy range 1-10 keV where the phenomenon is most visible in the 
analyzed data. 
These results contradict the common belief according to which the interactions 
between matter and antimatter are dominated by the 
reciprocally destructive phenomenon of annihilation. 
%A consequence of our results is that for energy 10-100 eV the antiproton
%is the ion with the best survival properties in the reflection process on a solid surface,
%since it does not capture electrons. 
%On the other side, the point where it comes to rest may 
%be known with  high precision.  Potentially, this makes it an exceptional 
%instrument to investigate small-scale 3-D complex structures 
%(like micro-cavities, or localized contaminations) and electromagnetic field small-scale layouts. 

\end{abstract}

%%%\pacs{36.10.-k, 34.80.Lx, 52.20.Hv}
\pacs{36.10.-k, 68.49.Sf, 13.40.-f}

\maketitle

\section{Introduction}

A large part of the experimentation with antiprotons after the 80's
has been devoted to 
the low energy interactions and properties of the antimatter-matter 
systems \cite{walcher88}. Among the other ones, this has produced relevant results concerning 
the $\bar{\mathrm H}$ production \cite{amoretti02,gabrielse02} and 
the properties of antimatter-matter systems on an atomic or 
molecular scale \cite{yamazaki93,zurlo06,elr07}. 
In this paper we present the experimental evidence of the reflection of a conspicuous 
fraction (20-30\%) of a beam of low energy (magnitude 1-10 keV) antiprotons hitting a solid 
surface. This contradicts the common belief according to which the interactions between antiprotons 
and matter at low energies are dominated by annihilations. 
The present analysis can be considered a refinement of the data analysis we presented 
in \cite{bianconi04}. Although the data were presented there, only 
a part of them was explained in that work.  

The experimental evidence refers to antiprotons that are reflected with  
energy $\sim$ a few keV, by a wall of solid aluminum. At these energies, 
the simulation of the reflection 
process shows that it is dominated by multiple Rutherford-like ``large angle'' 
scattering, where ``large'' means some tenths degrees. 
According to our simulation, 
the reflected fraction should increases at decreasing energy, possibly 
reaching 50\% at 500 eV.  

One century ago E. Rutherford wrote \cite{ruth} about the ``diffuse reflection'' of $\alpha$ and $\beta$ 
particles by thin metal layers, observing that the electromagnetic aspects of the process did 
not depend on the charge sign of the colliding particles. 
In the $\bar{p}-$nucleus case however the Rutherford mechanism competes with the annihilation process. 
At a different mass scale, it has been shown that 
positrons implanted into a variety of metals may 
return to the surface where they are re-emitted into the vacuum, possibly after capturing 
electrons and forming complex structures (see \cite{mills78,cassidy07} and references therein). 
Such processes may occur 
if the diffusion length is longer than the implantation depth. A similar situation takes place in 
the case considered here, where at the relevant energies 1-10 keV both the stopping range and 
the annihilation free path of the $\bar{p}$ in aluminum 
are longer than the path needed to lose memory of the initial 
flight direction. 

\section{This measurement}

\begin{figure}
\epsfig{file=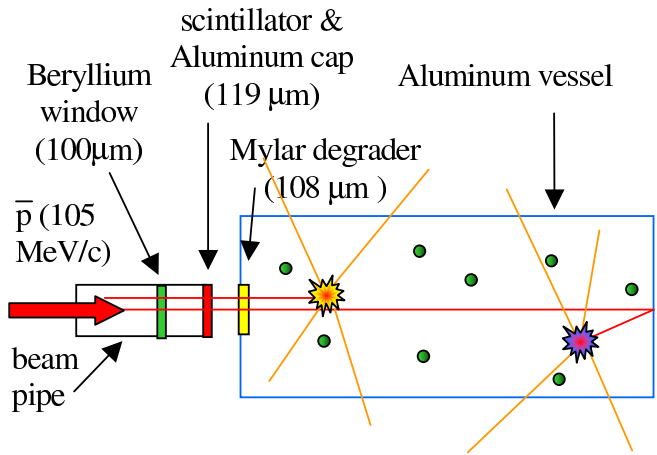,width=\columnwidth}
\caption{(Color) Layout of the beam line with antiprotons annihilation events in the target vessel.  
}
\label{figura1} 
\end{figure}

The data considered here belong to a set  of measurements performed at the LEAR 
(Low Energy Antiproton Ring) decelerator at CERN within the PS201 (OBELIX) experiment \cite{adamo92}. 
In these measurements, an antiproton beam with continuous energy distribution in the 
range 0-3 MeV enters a 75 cm-long  aluminum vessel containing 
the gas target (Fig.\ref{figura1}, see ref.\cite{zenoni00} for details on the apparatus). 
%Either in the gas itself, or on the vessel walls, 
%these antiprotons annihilated on some nucleus. 
As any charged particle, 
an antiproton that is travelling in matter loses progressively energy because of the 
stopping power of the crossed medium. 
It may rarely annihilate in flight, but in most cases the annihilation will take place when 
it is almost at rest, after the antiproton has been captured by an atom (see 
ref.\cite{zenoni00} for the separation of in-flight and at-rest annihilations). 
Antiprotons with entrance energy $\lesssim$ 4 keV 
come to rest and annihilate in the gas before reaching the end wall. 
Annihilation products (mesons) reach instantaneously the detectors out of 
the apparatus, and track interpolation allows for a precise reconstruction 
of the annihilation position (within cm) and time (within~ns).  
So, the $\bar{p}$ annihilations can be used to extract the main properties of the interactions between 
antiprotons and low pressure gases.
In such way it has been possible to determine e.g. the $\bar{p}$ stopping power in H$_2$,
 D$_2$ and He gases
down to capture \cite{bianconi04,bertin93,agnello95,elr02,elr04}. 
%Depending on which subset of these 
%data is considered, the kind of information that is extracted can be rather different, 
%ranging from properties of the in-flight annihilation process itself to the study of 
%``softer'' properties of the matter-antimatter interaction. 

\begin{figure}
\epsfig{file=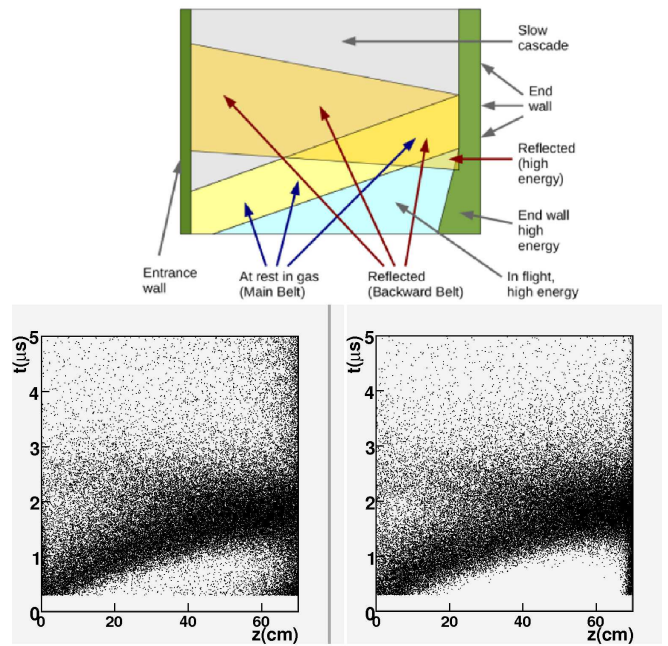,width=\columnwidth}
\caption{(Color) Top: scheme of the annihilation points of the data scatter plot (see text for explanation). 
Bottom left: experimental $(z,t)$-scatter plot. 
Bottom right: the simulated $(z,t)$-scatter plot.}
\label{figura2} 
\end{figure}

The data sample of the antiproton annihilations in helium at 1 mbar 
presented a puzzling feature 
left unexplained \cite{bianconi04} up to now. This feature is visible in the ``projected-path vs time'' 
scatter plot reported in Fig.\ref{figura2}, bottom left. 
Here, each point reproduces the longitudinal coordinate $z$ and the time of an annihilation event. 
Fig.\ref{figura2}, top, pictures the composition of the scatter plot population. 
%In fig.\ref{figura1}c each point of the scatter plot reproduces the longitudinal coordinate $z$ and the time
%$t$ of an annihilation event. In the considered case, the entrance wall is in position $z$=0,
%and the aluminum end wall at $z$=75 cm. 
The structure containing most of the annihilations in the gas is the so-called Main Belt, due to antiprotons 
entering the vessel  with energy below 3.5-4 keV, and coming to rest in gas before reaching the 
end wall. 
Initially, these particles slow down regularly because of the electronic stopping power \cite{agnello95}. 
At smaller than 0.5 keV 
energies, trajectory shape and energy loss are dominated by short-distance Rutherford collisions 
with helium nuclei and become irregular (nuclear stopping power). 
According to the Rutherford law the collision probability is proportional 
to  $\frac{1}{E^2 \sin^4(\frac{\theta}{2})}$ where $E$ is the kinetic energy of the antiproton and  
the scattering angle $\theta$ in the center-of-mass system,  
so large angle collisions are more likely at lower energies.  
Eventually, at energy below 30 eV, antiprotons are 
captured by helium atoms and form exotic atoms \cite{yamazaki93} with large quantum numbers. 
These systems undergo a statistical 
cascade processes leading to low-energy atomic levels within a time that depends on the density of the surrounding 
medium and is of the order of hundreds nanoseconds in helium at 1 mbar.
%When the scatter plot fig.\ref{figura1}c was published 
%first \cite{11} it was accompanied by a 

In our previous analysis \cite{bianconi04} the Monte Carlo simulation accounted for the 
just described processes (i.e. interactions with gas only), and reproduced the lower edge of the Main Belt. 
It was however unable to reproduce the 20-30\% fraction of annihilation points 
forming the ``Backward Belt'', i.e. the large secondary 
structure depicted in Fig.\ref{figura2} (top) and evident in the data of 
Fig.\ref{figura2} bottom left. 

\section{Simulated reconstruction of the data set}

Recently, the Monte Carlo simulation code has been improved, including the path of the antiprotons inside 
the aluminum end wall.  
To reproduce the behavior of the antiprotons in aluminum we have used standard atomic and metal parameters, 
plus an electronic stopping power that is about a half of the corresponding function for protons in aluminum, 
and is very close to the one measured by ref.\cite{moeller04}  
(we have fine-tuned it on our data; for a general discussion 
of the behavior of charged ions in matter at low energies see 
ref.\cite{sigmund00}). 
Below 1 keV there are no measurements of the antiproton behavior in aluminum, 
and we may only extrapolate the higher energy behavior 
%%new
(from a theoretical point of view interactions in the lower side of this region, where 
$E$ $\lesssim$ 100 eV, are extremely 
complex, since atomic and molecular degrees of freedom play an essential role. 
See \cite{Morgan88} and \cite{Cohen97} for discussions of these points).  
It must be remarked that the energy region below 1 keV (in aluminum) 
plays a marginal role here, 
since antiprotons that emerge from the end wall need to cover about 20-25 cm (back in the gas) 
to reach a position where their annihilation may be considered a signal of their reflection.  

For the electronic stopping power we use  $\frac{dE}{dx} = \alpha \times \sqrt{E}$ 
with $\alpha$  normalized by $\frac{dE}{dx} = \frac{360 \mathrm{MeV}}{\mathrm{g}/\mathrm{cm}^2}$ at $E$ = 1 keV. 
At a qualitative level these values are not critical, but they correspond 
to the best data   reproduction we have tested. 
The scattering between antiprotons and (helium or aluminum) atoms is treated 
by means of a screened Rutherford potential. The screening radius of the atomic cloud is 
$r_s$=1.25~\AA. Moderate
changes of $r_S$ have negligible effect on the results,
since the relevant collisions take place at impact 
parameters $\ll r_s$. 

In the vessel, the simulated antiprotons collide on helium atoms with a probability given 
by the total cross section $\pi r_s^2$. 
The electronic stopping power affects in a regular 
way the trajectory between two collisions, and is reproduced by the fit \cite{agnello95} 
$\frac{dE}{dx} = \alpha \times E^\beta$ where, for $dx$ in g/cm$^2$, $\alpha$ and $\beta$ 
are almost pressure-independent near 1 mbar; $\beta$ $=$ 0.29 and $\alpha$ is fixed so to have 
$dE/dx$ $=$ 35 eV/cm at 1 keV and 1 mbar. These 
values have been first extracted in \cite{agnello95} at 4 and 8.2 mbar, and later 
confirmed in \cite{bianconi04} at 1 and 0.2 mbar. 
The angle and energy loss at each collision is determined 
by screened Rutherford scattering. The only involved parameter is $r_s$ $=$ 
$1.5 r_b/2$ ($r_b/2$ is the Bohr radius for He$^+$). 
The values of the helium parameters $r_s,\alpha,\beta$ 
are fixed by the measurement at 0.2 mbar in \cite{bianconi04} in a way that is not affected by the 
reflection process discussed here, since 
in the 0.2 mbar case the vessel is much longer and the end wall has no effect on the examined 
vessel region. The cascade time has been best-fitted to 0.4 $\mu s$ 
in the present work. 

\begin{figure}
\epsfig{file=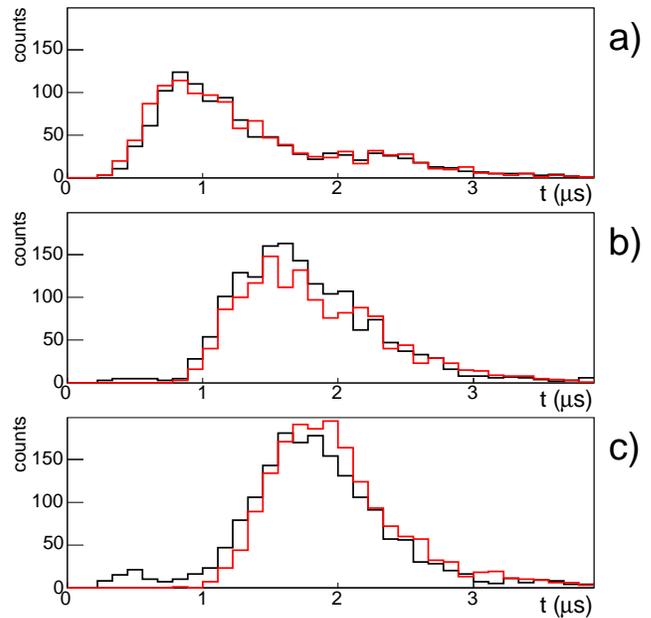,width=\columnwidth}
\caption{(Color) Comparison between experimental (black) 
and Monte Carlo (red) data. 
Time-distribution of the subsets of events 
corresponding to three different slices in $z$ coordinate: (a): 14 cm $< z <$ 16 cm; (b):  40 cm $< z <$ 42 cm; 
(c): 56 cm $< z <$ 58 cm. }
\label{figura3}
\end{figure}

For the simulation of Fig. \ref{figura2} (bottom right) and Fig.\ref{figura3} we have considered 270,000 
antiprotons entering the vessel with energy homogeneously distributed between 0 and 30 keV.  Some 
preliminary 
simulations have been pushed to 80 keV, showing that particles between 30 and 80 keV do not introduce 
relevant differences in the regions of the Main and Backward Belt.  

Because of the wide energy spectrum extending up to 3 MeV, the very largest 
part of the annihilation events 
occurs on the end wall after some tens of nanoseconds (time of flight). 
For this reason we have removed from the present data analysis the 
antiproton annihilations occurring in the first 250 ns.
At larger times, initial-wall and end-wall annihilations 
affect the regions $z$ $<$ 5 cm and $z$ $>$ 60 cm (green areas in Fig.\ref{figura2}, top)  
because of the 1 cm Gaussian uncertainty 
in the measurement of $z$, and of the huge number of these annihilations. 

The scatter plot of Fig.\ref{figura2} (bottom right) has been simulated  
including propagation and multiple scattering inside the end wall, 
and it can be considered satisfactory. This is not only evident 
from the comparison of experimental and simulated scatter plots (bottom panels of Fig.\ref{figura2}),  
but also from the comparison of the time distributions 
for events with given $z$. This is shown in Fig.\ref{figura3}, where we 
report three subsets of the events of the scatter plots of Fig.\ref{figura2}. 
These subsets correspond to (top) 14 cm $< z <$ 16 cm,  
(middle) 40 cm $< z <$ 42 cm,  (bottom)  56 cm $< z <$ 58 cm. The time distributions of 
these subsets are reported for 
the experimental and for the simulated events (black 
and red histograms in Fig.\ref{figura3}). Within standard statistical fluctuations,
 the simulation reproduces the shape and also the normalization of the experimental 
 distributions.  

The present reproduction of the data may be compared with the one of \cite{bianconi04}. The difference 
introduced by including the effect of the end wall is evident, both in the scatter plot and in the 
$z-$slices. 

%The average lifetime of the bound $\bar p - he$ atom depends on the density in a not obvious 
%way and must be fitted at any given density. 
%In the case of the measurement at 1 mb, with end wall 
%at $z$ $=$ 75 cm, if one neglects the effect of 
%the reflection on the end wall the $z-$averaged lifetime is artificially enlarged by reflection-originated 
%annihilations at large t. For this reason, here we have best-fitted 0.4 �s, that is shorter than the 
%time fitted in ref.\cite{11}. Values between 0.3 and 0.5 $\mu$s are good as well. 

\begin{figure}
\epsfig{file=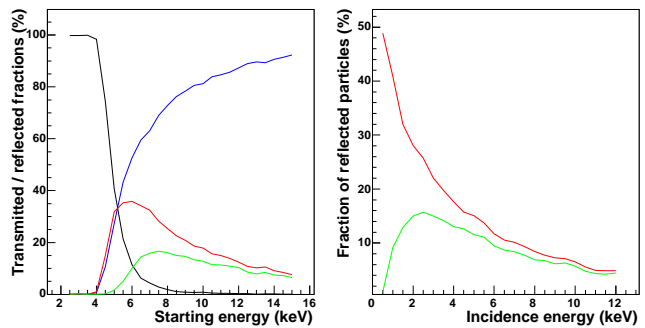,width=\columnwidth}
\caption{(Color) Simulated distributions. Left: distribution with respect 
to the energy at the entrance of the gas container, 
for antiprotons that come to rest in gas before reaching the end wall (black), 
that reach the end wall (blue), that are reflected by the end wall (red), that are reflected 
with enough energy to reach $z <$ 65 cm (green). Right: distribution 
with respect to the energy when hitting the aluminum wall, for antiprotons that are reflected (red), that are 
reflected 
with enough energy to reach $z <$ 65 cm (green).  }
\label{figura4}
\end{figure}

\section{Discussion}

\begin{figure}
\epsfig{file=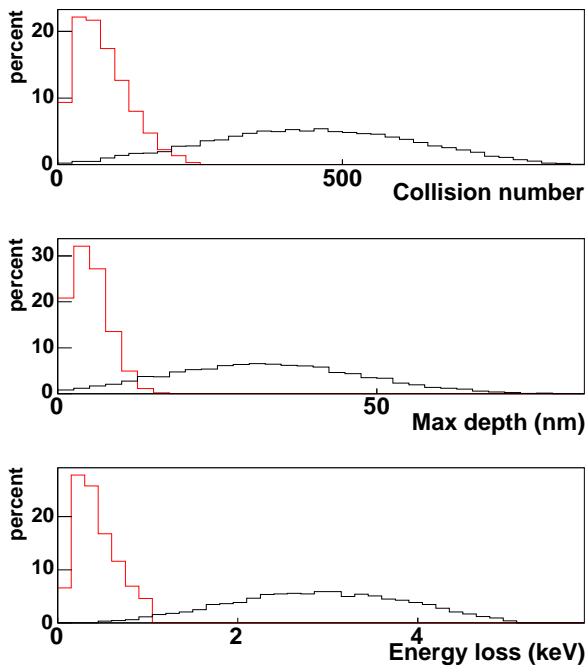,width=\columnwidth}
\caption{(Color) Relevant distributions (in percent) for in-wall  
features of antiprotons: in red for 1 keV energy,  
in black for 5 keV energy. These distributions only refer to antiprotons 
that are reflected. See text for details.  }
\label{figura5}
\end{figure}

In Figs.\ref{figura4} and \ref{figura5} the simulation code is used to study some relevant 
properties of the reflection process. 
In Fig.\ref{figura4} (left) we compare the dependence on the initial energy  
(the energy of the antiprotons when they enter 
the vessel) of the fraction of particles that (i) stop in helium before reaching the end wall, (ii) come to 
rest in aluminum, (iii) are reflected, (iv) are reflected with enough energy (about 500 eV) to get 10 cm backward 
in the gas at least. In Fig.\ref{figura4} (right) the distributions (iii) and (iv) 
are considered as functions of the energy owned by the 
antiprotons when they hit the aluminum wall with normal incidence. Fig.\ref{figura5} is devoted 
to the behavior of the antiprotons 
inside aluminum. The red histograms of Fig.\ref{figura5} refer to 20,000 
antiprotons entering aluminum with normal incidence 
and energy 1 keV. Of these, 7,487 are reflected. For these particles only, we show the 
distribution (in percentage terms) of the number of scattering events inside aluminum, 
of the maximum longitudinal depth reached inside the wall, and of the lost energy. 
The same distributions are reported in the black histograms of Fig.\ref{figura5} for 11,347 reflected $\bar{p}$
out of 80,000 ones with energy 5 keV. 

In Fig.\ref{figura5} ``scattering event'' means any event where the 
antiproton passes within 1.25 \AA~from the target nucleus. 
At all the energies where we have measurements of the annihilation cross section of antiprotons on some nucleus, 
this is many orders of magnitude smaller than the atomic-size cross section for elastic scattering. 
Decreasing the energy, the annihilation cross sections at low energy should follow 
a ``1/energy'' law, that is typical for inelastic processes between hadrons 
with opposite electric charge (see e.g. \cite{bianconi00} and references therein). 
So we need to check that annihilation cross sections 
do not grow to a size such as to obscure elastic reflection effects. 
For hydrogen, deuterium and helium targets, where we have measurements  
at very small energies \cite{zenoni99,bertin96}, we have 
$\sigma_{ann} \approx \frac{c^2}{v^2} \times 10^{-27}$ cm$^2$, 
where $v$ is the antiproton velocity and $c$ is the speed of light.  
Assuming that in the aluminum case the annihilation cross section is $N$ times larger than in the helium or 
hydrogen cases, 
and taking $\sigma_{el}$ $=$ (1.25\AA)$^2$, 
we have $\sigma_{ann}/\sigma_{el}$ $\approx$ $2\cdot 10^{-6}N/E$, for $E$ expressed in keV. 
No data or theory gives us elements to imagine 
that $N$ may overcome 100 in magnitude. This means that  both at 1 keV 
 and 5 keV the number of scattering events reported in Fig.\ref{figura5} corresponds to a path that is much 
shorter than the average annihilation path. 

The physics emerging from these data is simple: at the distance scale of the nuclear radius it is 
fair to consider antiprotons as ``destructive'' particles, but for an antiproton with energy 
1-10 keV the probability 
of finding itself within such a distance from the nucleus is several orders of magnitude smaller than the 
probability of being deflected by an angle 20 degrees. After 10 such collisions the memory of the initial 
direction would be completely lost, and 50\% of the antiprotons would be backward directed. 
At energy 5 keV we do not have so many large angle collisions, but their number is anyway large enough 
to allow a relevant fraction of the antiprotons to be reflected. 
The Monte Carlo simulation shows that $multiple$ scattering (with angles $10^\circ-40^\circ$)  
dominates the reflection process 
at the energies that are central here (some keV). At energies $<$ 1 keV single scattering with 
angle $>$ 90$^\circ$ becomes relevant too. 

%The most promising applications of the antiproton reflection are related with a 
%combination of two relevant specificities,
%with respect to positive ions (for the application of the reflection of low energy positive ions to surface 
%analysis see e.g. \cite{niehus93,kasi89}). 
%First, positive ions capture electrons and become invisible at energies of magnitude 100 eV.
%The antiproton cannot be made charge-neutral until it 
%reaches capture energy (magnitude 10 eV). So, at very low energies 
%the antiproton is the ``best-survival'' ion probe 
%potentially available, with electromagnetic fields allowing for driving its
%trajectory up to capture energy. Second, when antiprotons are captured the annihilation follows promptly, and the
%annihilation vertex position and time may be reconstructed with great precision. On a smaller space scale, this 
%is the analog of the Positron Emission Tomography 
%method \cite{phelps75}: in both cases annihilation products allow for a
%precise reconstruction of the annihilating particle position/time. We may so imagine scenarios where these
%particles are used to explore microscopic cavities, arrangements of electric fields, localized impurities,
%and so on.  

In the near future, the systematic experimental study of antimatter reflection can be realized at AD at CERN,
for example by the ASACUSA Collaboration \cite{hayano08}, or at the future 
low energy antiproton facility FLAIR at GSI \cite{flair}.
 

\begin{thebibliography}{99}


\bibitem{walcher88} T. Walcher, Ann. Rev. Nucl. Part. Sci. {\bf 38}, 67 (1988).
%\bibitem{baird96} S. Baird et al., CERN/PS 96-43 (1996)

\bibitem{amoretti02} M. Amoretti {\it et al.},  Nature {\bf 419}, 456 (2002).

\bibitem{gabrielse02} G. Gabrielse {\it et al.}, Phys.\ Rev.\ Lett.\ {\bf 89}, 213401 (2002).

\bibitem{yamazaki93} T. Yamazaki {\it et al.},  Nature {\bf 361}, 238 (1993). 

\bibitem{zurlo06} N. Zurlo {\it et al.}, Phys.\ Rev.\ Lett.\ {\bf 97}, 153401 (2006).

\bibitem{elr07} E. Lodi Rizzini, L. Venturelli \& N. Zurlo, ChemPhysChem {\bf 8}, 1145 (2007). 

\bibitem{bianconi04} A. Bianconi {\it et al.}, Phys.\ Rev.\ A {\bf 70}, 032501 (2004).

\bibitem{ruth} E. Rutherford, Philos.\ Mag. {\bf 21}, 669 (1911).

\bibitem{mills78} A.P. Mills Jr., Phys.\ Rev.\ Lett. {\bf 41}, 1828 (1978). 

\bibitem{cassidy07} D.B. Cassidy \& A. P. Mills Jr., Nature {\bf 449}, 195 (2007).

%\bibitem{niehus93} H. Niehus, W. Heiland \& E. Taglauer, Surf.\ Sci.\ Rep.\ {\bf 17}, 213 (1993).

%\bibitem{kasi89} S.R. Kasi {\it et al.},  Surf.\ Sci.\ Rep.\ {\bf 10}, 1 (1989).

%\bibitem{armenteros86} R. Armenteros et al., CERN-PSCC-86-4; PSCC-P-95 (1986) 

\bibitem{adamo92} A. Adamo. {\it et al.},  Sov.\ J.\ Nucl.\ Phys.\ {\bf 55}, 1732 (1992).

\bibitem{zenoni00} A. Zenoni. {\it et al.},  Nucl.\ Instr.\ and Meth.\ A {\bf 447},  512 (2000).

\bibitem{bertin93} A. Bertin {\it et al.}, Phys.\ Rev.\ A {\bf 54} 5441 (1993).

\bibitem{agnello95} M. Agnello {\it et al.}, Phys.\ Rev.\ Lett.\ {\bf 74} 371 (1995).

\bibitem{elr02} E. Lodi Rizzini {\it et al.}, Phys.\ Rev.\ Lett.\ {\bf 89}, 183201 (2002).

\bibitem{elr04} E. Lodi Rizzini {\it et al.}, Phys.\ Lett.\ B {\bf 599}, 190 (2004).

\bibitem{moeller04} S.P. M\o ller {\it et al.}, Phys.\ Rev.\ Lett.\ {\bf 93}, 042502 (2004).

\bibitem{sigmund00} P. Sigmund \&  A. Schinner, Eur.\ Phys.\ J.\ D  {\bf 12}, 425 (2000). 

\bibitem{Morgan88} D.L. Morgan Jr, Hyperf. Int. {\bf 44} 399 (1988).

\bibitem{Cohen97} J.S. Cohen, Phys. Rev. A {\bf 56}, 3583 (1997). 

\bibitem{bianconi00} A. Bianconi {\it et al.}, Phys. Lett. B {\bf 483},  353 (2000).

\bibitem{zenoni99} A. Zenoni {\it et al.}, Phys.\ Lett.\ B {\bf 461}, 405 (1999). 

\bibitem{bertin96} A. Bertin {\it et al.}, Phys.\ Lett.\ B  {\bf 369}, 77-85 (1996).

%\bibitem{phelps75} M.E. Phelps, E.J. Hoffman, N.A. Mullani \&
%M.M. Ter-Pogossian, J.\ Nucl.\ Med.\ {\bf 16}, 210 (1975).

\bibitem{hayano08} R.S. Hayano {\it et al.}, SPSC-SR-027 ; CERN-SPSC-2008-002 (2008).

\bibitem{flair} http://www.oeaw.ac.at/smi/flair/


\end{thebibliography}

\end{document}